
\input harvmac

\def \Pp{{\bf P}}

\Title{\vbox{\hbox{HUTP--95/A035}\hbox{IASSNS--HEP--95--77}}}
{\vbox{\centerline{D-Strings on D-Manifolds}}}
\vskip2ex
\centerline{M. Bershadsky and C. Vafa}
\vskip2ex
\centerline{\it  Lyman Laboratory of Physics, Harvard
University}
\centerline{\it Cambridge, MA 02138, USA}
\vskip2ex
\centerline{and}
\vskip2ex
\centerline{V. Sadov}
\vskip2ex
\centerline{\it School of Natural Sciences and  School of
Mathematics, }
\centerline{\it Institute for Advanced Study}
\centerline{\it Olden Lane, Princeton, NJ 08540, USA}
\vskip .3in
We study the mechanism for appearance of massless solitons
in type II string compactifications.  We find that by combining
$T$-duality with strong/weak duality of type IIB in 10 dimensions
enhanced gauge symmetries and massless solitonic hypermultiplets
encountered in Calabi-Yau compactifications
can be studied perturbatively using D-strings (the strong/weak dual to
type IIB string) compactified on ``D-manifolds''.  In particular
the nearly massless solitonic states of the type IIB compactifications
correspond to elementary states of D-strings.  As examples we consider
the D-string description of enhanced gauge symmetries
for type IIA string compactification on
ALE spaces with $A_n$ singularities and type IIB on a class of singular
Calabi-Yau threefolds.  The class we study includes as a special case
the conifold singularity in which case the perturbative
spectrum of the D-string includes the expected massless
hypermultiplet with degeneracy one.

\input epsf

\Date{10/95}

\newsec{Introduction}
One of the major recent discoveries in string theory
has been the appreciation of the existence and importance
of solitonic objects
\ref\hu{C. M. Hull and P. Townsend, Nucl. Phys. {\bf B438}  (1995) 109}
\ref\Wt{E. Witten, Nucl. Phys. {\bf B443}  (1995) 85}
\ref\st{A. Strominger, Nucl. Phys. {\bf B451}  (1995) 96}
\ref \pol{J. Polchinski, {\it Dirichlet-Branes and Ramond-Ramond
Charges},
Preprint NSF-ITP-95-122, hep-th/951001}.
In particular
it has become clear that in the case of type IIA compactifications on
$K3$ when we have a vanishing 2-cycle we get enhanced gauge
symmetries.
In case of  type IIB compactification on Calabi-Yau threefolds
vanishing three-cycles have been proposed to lead to massless charged
matter.  The basic mechanism suggested \st\
is the appearance of solitons corresponding
to wrapping of $p$-branes around  vanishing cycles.
The exact conformal theory describing
the coupling of such solitons to string theory is recently proposed in
\pol\  as D-branes. Even with such a nice picture emerging, we still lack
a perturbation scheme  in which the wrapped p-branes
appear as fundamental string states specifically for $p>1$, which
is the case typically encountered.  In this paper
we will argue that, at least for all the cases encountered
thus far,
by using $T$-duality (which leads to backgrounds with
$H$-fields turned on) the value of $p$ can be changed to $p=1$.
Since the $p=1$ D-branes are simply the D-strings,
 dual to type IIB string under strong/weak duality,
the nearly massless
states are expected to show up in the perturbative spectrum
of D-string propagating on the dual compactification
which we will call the `D-manifold'.
`D-manifolds' are manifolds with D-brane configurations
(`skeletons').
  We will find that this is indeed the case.

The main motivation behind this
work was to better understand the appearance of massless
modes in type II string compactifications on Calabi-Yau manifolds.
We will first consider type IIA compactification on
$K3$ and study how the gauge symmetry enhancement is expected
to arise.  We find similarity with how the gauge
symmetry is enhanced on the $D$-brane worldsheet
through the appearance
of massless states in the open string sector
when $D$-branes approach each other
\ref\polt{J. Polchinski,
Phys. Rev. {\bf D50} (1994) 6041}
\ref\ed{E. Witten, {\it Bound states of Strings and p-Branes},
IASSNS-HEP-95-83 Preprint, hep-th/9510135} \ref\polwit{J.
Polchinski and E.
Witten,
{\it Evidence for Heterotic Type I Duality, Preprint  IASSNS-HEP-95-81,
hep-th/9510169}}.  This similarity is
not an accident and we argue that 10 dimensional
strong/weak duality of type IIB maps one to another.  In particular
combining type IIB strong/weak duality
with the fact that coincident $D$-branes naturally give
rise to enhanced
gauge symmetry through standard Chan-Paton factors
{\it explains} the appearance of gauge
symmetry for special moduli of type IIA compactification on $K3$.
More precisely we will find that type IIA on ALE space
with $A_{n-1}$ singularity is equivalent to D-strings on
a D-manifold which contains $n$ parallel, coincident
Dirichlet 5-branes.

We then consider the compactifications of type IIB
on singular Calabi-Yau threefolds.  The simplest
type is where the Calabi-Yau develops a conifold singularity for which
there is a prediction \st\ for the
appearance of a massless hypermultiplet.  To gain
a better insight it is natural to consider more general
singularity types.  One class is suggested by the observations in
\ref\ghv{D. Ghoshal and  C. Vafa, {\it C=1 Strings as
the Topological Theory of the Conifold},
Preprint HUTP-95-A022, hep-th/9506122}\
that the conifold
singularities are topologically equivalent to
$c=1$ strings at the self-dual radius coupled to gravity
 (by making use of the identification of $c=1$ at the
self-dual
radius with the twisted Kazama-Suzuki coset $SL(2)/U(1)$ at $k=3$
\ref\mv{S. Mukhi and C. Vafa, Nucl. Phys. {\bf B407}  (1993) 667}).
One evidence in favor of this was the fact that the coordinate
ring of the conifold agrees with the ground
ring of the $c=1$ at the self-dual radius
\ref\zw{E. Witten, Nucl. Phys. {\bf B307}  (1992) 187;
E. Witten and B. Zwiebach, Nucl. Phys.  {\bf B377}  (1992) 55}
\foot{  To relate it to Calabi-Yau compactification,
non-critical
$c=1$ strings compute the leading behavior for prepotential and some
special amplitudes involving $(2g-2)$ graviphotons and $2$ gravitons
\ref\bcov{M. Bershadsky, S. Cecotti, H. Ooguri and C. Vafa,
Comm. Math. Phys. {\bf 2}  (1994) 311}
\ref\narainet{I. Antoniadis, E. Gava, K.S. Narain, T.R. Taylor, Nucl.
Phys. {\bf B413}  (1994) 162}.
The computation of \ref\nare{ I. Antoniadis, E. Gava, K.S. Narain
and T.R.
Taylor,
{\it N=2 Type II Heterotic Duality and Higher Derivative F Terms},
Preprint IC-95-177, hep-th/9507115}\
for such corrections in the presence of a nearly massless hypermultiplet
and its agreement with the partition function of $c=1$ at the self-dual
radius
provides a strong all genus check for the appearance of the massless
hypermultiplet at the conifold points of Calabi-Yau moduli \st.}.
Given the rich physical content of $c=1$ theories and its direct
implications for physics of type II compactifications on Calabi-Yau
manifolds,
it is natural to wonder what type of physics emerges when we consider
compactification
on Calabi-Yau manifolds which have a singularity
corresponding to ground ring of other
 $d=2$ strings, classified by $ADE$.

We will mainly concentrate on the A-type $d=2$ strings, i.e.
when the Calabi-Yau develops a singularity corresponding to the
ground ring of $2d$ string for $c=1$ at $n$-times the self-dual
radius
\ref\mukgh{D. Ghoshal, D. P. Jatkar and
S. Mukhi, Nucl.Phys. {\bf B395}  (1993) 144}.  We find that for
$n$-times the self-dual radius there is a $U(n)$ enhanced gauge
symmetry as well as matter in the fundamental and adjoint
representations
of $U(n)$.  The mechanism for this rich structure
of massless states is having different
types of vanishing three cycles where the three cycle collapses
to a point or to a curve
(this point has also
been noted independently in
\ref\aspn{P. S.  Aspinwall, {\it N=2 Dual Pair and a Phase Transition,
Preprint  CLNS-95-1366, hep-th/9510142}}).

Following the connection with D-branes in the case
of $K3$ compactifications we find the
$D$-string duals
for the singular limits of the corresponding
Calabi-Yau.  In particular the
 relevant D-manifold consists of $n+1$ 5-branes,
where $n$ of them are parallel to one another and intersect
the other 5-brane on the 3+1-dimensional space-time.  This allows
us to have a simple description for the degeneracy of
nearly massless hypermultiplets.  As a special case we
derive the fact that in the case of the conifold singularity
there is exactly one massless hypermultiplet, as had
been proposed \st .

\newsec{K3 and Gauge Symmetry Enhancement}

It is interesting to study in detail how type IIA on $K3$
leads to gauge symmetry enhancement through vanishing
2-cycles.  For simplicity we will mainly consider  $A_{n-1}$
singularities
but make some comments about $D$ and $E$ as well.
For our purposes, the {\it local } analysis near singularity is
sufficient. Degenerating family of  $K3$'s can locally be modelled by
a
family of (resolved) $A_{n-1}$ ALE spaces:  surfaces  given by
\eqn\reso{\prod_{i=1}^{n}(\zeta -\mu_i)+z^2+w^2=0.}
in ${\bf C}^3$ with coordinates $z$, $w$ and $\zeta$.  This family is
parameterized by   $\{\mu_i\}_{i=1}^n$ that may
serve as  local coordinates on moduli space of $K3$.
To be precise we also have to specify a point on the K\"ahler moduli
for $K3$.
 For all
$\mu_i\neq
\mu_j$,  the surface \reso\ is smooth.  When $\mu_i$ meets $\mu_j$,
an
$A_1$ singularity develops and a 2-sphere  $S^2_{ij}$ vanishes.
If we take $\mu_i$ to be real the
 vanishing 2-sphere can be viewed as a real section of \reso .
Similarly, when $\mu_{i_1}=\ldots =\mu_{i_k}$ the surface \reso\
develops
$A_{k-1}$ singularity with $k-1$ independent cycles
$S^2_{i_1i_2},\ldots,
S^2_{i_{k-1}i_k}$ shrinking to zero size.

\bigskip

\let\picnaturalsize=N
\def\picsize{2.0in}
\def\picfilename{ren1.eps}
\ifx\nopictures Y\else{\ifx\epsfloaded Y\else\fi
\global\let\epsfloaded=Y
\centerline{\ifx\picnaturalsize N\epsfxsize \picsize\fi
\epsfbox{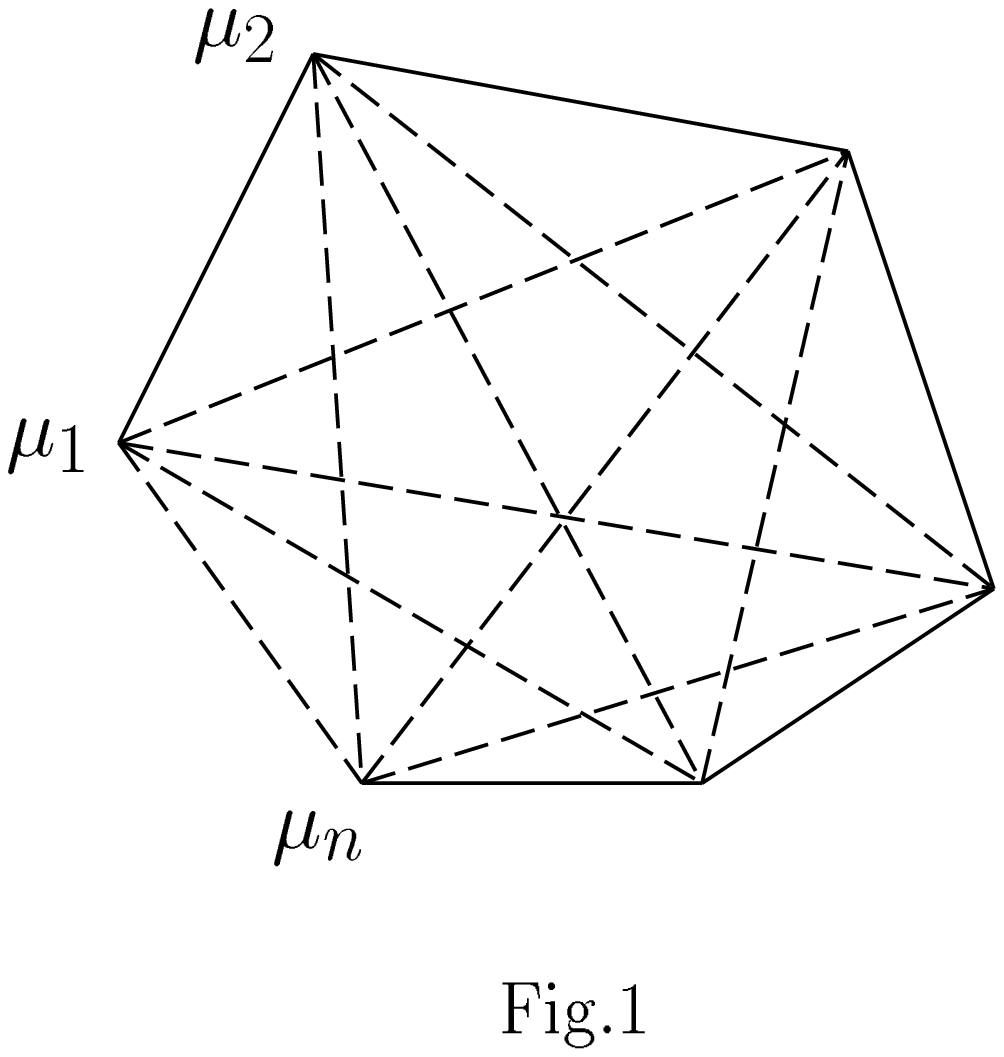}}}\fi

\bigskip

Figure 1 captures the $\mu$-plane.  Each line connecting
 two values of $\mu_i$ represents a vanishing $S^2$.
A sublattice in $H_2(K3)$
generated by all cycles $S^2_{12},\ldots, S^2_{(n-1) n}$  coincides
with
$A_{n-1}$ root lattice.
The monodromy of a cycle $S$ around a singular locus $\mu_i=\mu_j$
is
given by
Picard-Lefshets formula
\eqn\piclef{S \rightarrow S- S_{ij} (S \circ S_{ij}).}
The intersection form on  vanishing 2-cycles coincides with Cartan
form
on  $A_{n-1}$ and therefore the monodromy
\piclef\ is just a Weyl reflection.  The whole monodromy group
coincides
with a Weyl group of $A_{n-1}$. It is important that taking a
vanishing
cycle, say  $S^2_{12}$, and applying the monodromy group to it one
obtains
{\it all} other vanishing cycles.
In Type IIA compactification on $K3$ extra light multiplets
are expected to appear
in such a limit \Wt\ provided that
the $B$-field part of the K\"ahler moduli is suitably chosen
\ref\aspi{P. S. Aspinwall, Phys. Lett. 357{\bf B} (1995) 329}.
The basic mechanism \st\ (suggested in the context
of Calabi-Yau threefolds) is that there are nearly massless
solitons where
appropriate 2-branes wrap around the vanishing 2-cycles.  The precise
details of how this happens has not been clarified.   A promising
candidate of the corresponding 2-branes has been found recently
\pol\ to be a D-brane which has a simple conformal
theory description.

\subsec{D-branes and Gauge Symmetries}
It is quite striking that the mechanism of enhanced gauge
symmetry we observed here is very similar to the mechanism
of enhancement of gauge symmetry on the worldsheet of $D$-branes
when they approach each other \polt \ed \polwit.  In particular
the masses of massive gauge particles are measured by the separation of
pairs of points $\mu_i,\mu_j$
\foot{Actually note the the K\"ahler moduli
should also be added to the $\mu_i$ to give
a correct measure of mass as explained before.}.
In the case of the D-branes,
if we have $n$ parallel nearby $D$-branes, there is an enhancement
of gauge symmetry to $U(n)$.  The corresponding gauge particles
are given by open string states connecting two nearby D-branes.
Again the mass of each of these nearly massless gauge bosons is
related to
the stretching of open strings, or equivalently to how far they are
from one another.  We will now see that
this amazing similarity is not an
accident
and leads to an unexpected connection between type IIB strong/weak
duality in 10
dimensions and gauge symmetry enhancement for
type IIA upon compactification on $K3$ with $A_{n-1}$ singularity.

To understand this connection it turns out to be crucial
to study the structure of conformal theory near the singularities
of $K3$.  This has been done recently
\ref\sov{H. Ooguri, S. Shenker and C. Vafa, in preparation}\
and it turns out that the conformal theory is exactly solvable
for all the $ADE$ singularities of $K3$.  What is found is that
for the $A_{n-1}$ case it is (a capped version of) the same conformal
theory as that
of symmetric 5-branes
\ref\strhc{C. G. Callan, J. A. Harvey and A. Strominger,
Nucl. Phys. {\bf B367}, (1991) 60}\
with $H$ charge $n$. A connection between the conformal theory
associated to $A_1$ singularity and symmetric fivebrane
 had been anticipated
\ref\wit{E. Witten, {\it Some Comments on String Dynamics},
Preprint IASSNS-HEP-95-63, hep-th/9507121}.
However a crucial
subtlety appears leading to an exchange between type IIA and type IIB:
It turns out that
type IIA on ALE space with $A_{n-1}$ singularity is equivalent to
type IIB on
symmetric fivebranes of $H$-charge $n$.  The basic idea is
that if we think of $K3$ as fibered over the sphere with tori
as the fiber (where the torus becomes singular at the analog
of $\zeta =\mu_i$) the $R\rightarrow 1/R$ duality transformation of
the fiber
(mapping type IIA to type IIB)
implies that points where the torus degenerates are also the points
where there is an $H$ charge \ref\cosm{B. R. Greene, A. Shapere, C. Vafa
and S.-T. Yau, Nucl. Phys. {\bf B337} (1990) 1}.

In this context we thus expect that type IIB on symmetric fivebranes
leads to enhanced gauge symmetries in six dimensions.  What would
be a candidate for the corresponding soliton?  Being the dual to type
IIA description it can be either a 1-brane or a 3-brane.  We will
now see that they are indeed 1-branes as is suggested
by Figure 1.

The conjectured
$SL(2,Z)$ duality of type IIB in 10 dimensions
\hu\ has been recently analyzed
in connection with the prediction of new types of strings
\ref\schw{J. H. Schwarz {\it An SL(2,Z) Multiplet of Type IIB
Superstrings},
Preprint CALT-68-2013, hep-th/9508143}\
and in connection with the proposal
of Polchinski \pol\
in constructing bound states for D-branes \ed.
The strong/weak duality
of type IIB in 10 dimensions maps
the $B$ field of the NS-NS sector to the $\tilde B$ field of the
R-R sector. Strings are the source of $B$ and Dirichlet 1-branes,
the D-strings,
are the source of $\tilde B$.  Thus strong/weak duality
maps type IIB strings to D-strings.
To search for 1-brane solitons, which are also the same
as D-strings, all we need to do is
to study D-strings perturbatively where the
solitons would appear as elementary D-string states.
We have to construct what the background looks
like for the D-strings.
Since
the symmetric five branes are the source for the field
which is spacetime dual\foot{
The duality here means that the field strengths
are Poincare dual.} to $B$
and Dirichlet 5-branes are the source for the spacetime dual to
$\tilde B$
field, we see that strong/weak duality should exchange the
symmetric 5-branes with the Dirichlet 5-branes.  In other words
type IIB strings on a manifold consisting of n-symmetric 5-branes
is dual to D-strings on a D-manifold consisting of n parallel Dirichlet
5-branes.

On the other hand type IIB on a manifold with a
symmetric 5-branes with $H$ charge equal to $n$
is equivalent to type IIA
on a space with $A_{n-1}$ singularity \sov.
Note that on the 6 dimensional worldsheet of each of
 the Dirichlet 5-branes we have a
$U(1)$ gauge symmetry.  Moreover we expect that when
 $n$ Dirichlet 5-branes approach each other we obtain,
through the standard Chan-Paton factors,
an enhancement of gauge symmetry from $U(1)^n\rightarrow U(n)$,
i.e. a $U(n)$ gauge symmetry in 6 dimensions.
This gauge symmetry
enhancement is exactly what we would expect for type IIA on
ALE spaces with $A_{n-1}$
singularity!  (the extra $U(1)$ is related to the $U(1)$
gauge field of type IIA in 10 dimensions).  We have thus
connected type IIA enhancement of gauge symmetry
to type IIB strong/weak duality in 10 dimensions through standard
Chan-Paton mechanism.  Moreover the solitonic
Dirichlet 2-branes of type IIA which are becoming massless
are mapped to the nearly massless open string states of D-string
(depicted by lines connecting pairs of $\mu_i$ in Figure 1).
Note also that the fact that we have to adjust {\it four} real
quantities to get a massless states is consistent and even more
clear in the
D-string description.

\subsec{Type IIB on $K3\times T^2$}
In preparation for our discussion of the Calabi-Yau threefolds
we wish to discuss type IIB on $K3\times T^2$ (changing
from our starting point which was type IIA on $K3$).
Type IIB compactification
 on $K3\times T^2$ is equivalent to type IIA compactification
on $K3\times \tilde T^2$ where $\tilde T^2$ is the torus dual
to $T^2$ with K\"ahler and complex moduli exchanged.
We can now have a different perspective about the
appearance of massless particles when we consider
$K3$ with vanishing 2-cycles
\wit:
 The extra
massless gauge particles required by string duality can
be understood from the view point of the vanishing 3-cycles.
 The geometry of vanishing cycles $V_{ij}$
in this case is $S_{ij}^2\times S^1$, where the $S_{ij}^2$ comes from
the
singular locus on $K3$, where  $\mu_i$ meets $\mu_j$,
and $S^1$ is one of the cycles of the $T^2$.
Cartan subalgebra of $SU(n)$ corresponds to complex moduli $\mu_i
-\mu_{i+1}$,
while the nilpotent subalgebras ${\cal N}_{\pm}$ are spanned by
solitonic
configurations --
three-branes wrapped around vanishing cycles.  Primitive cycles
$V_{ii+1}$
correspond to simple roots
(solid lines on Figure 1), while  $V_{ij}$ are in one to one
correspondence with  positive non simple
root vectors (dashed lines on Figure 1). Negative roots correspond to
three-brane configuration with
reversed orientation.

Let us discuss the monodromy group for $K3\times T^2$. Since we do
not
vary a modulus of $T^2$, the parameter space is the
same as for $K3$. Thus the monodromy group is the Weyl group of
$A_{n-1}$
again.  An action of the Weyl group on 3-cycles is induced by its
action
on 2-cycles of $K3$. In particular, if we denote by $A$ and $B$ two
circles of $T^2$, the cycles $\{S^2_{ij}\times A\}$ are never mixed
with
the
cycles $\{S^2_{ij}\times B\}$. Moreover, cycles from the same group do
not
intersect\foot{The fact that
there is still a nontrivial monodromy does not contradict
Picard-Lefshets
simply because the setup is different.}:
$(S^2_{ij}\times A) \circ (S^2_{kl}\times A)=0$. In the context of
Type
IIB
compactification this means that in 4 dimensions, the particles
corresponding to $\{S^2_{ij}\times A\}$ are mutually local. On the
other
hand, since $S^2_{ij}\times A$ intersect $S^2_{kl}\times B$ the
full set of 4-dimensional particles is not mutually local. The
$U$-duality
transformation of $T^2$ exchanges $A$ and $B$, making
electric-magnetic
duality transparent \ed.

 The hypothesis  is that the solitonic configurations we just
described
  are the only stable ones.  This partially  follows from
transitivity of  monodromy group as it acts on vanishing cycles.
Indeed,
applying monodromies to a stable solitonic configuration
corresponding to
$V_{12}$ one gets {\it all} other solitons provided  one does not
encounter jumping phenomena, which is guaranteed by $N=4$
supersymmetry.
Note that since we have $N=4$ in $d=4$ we should expect in the
$N=2$ terminology one massless vector multiplet and one massless
hypermultiplet both in the adjoint of $SU(n)$.

It should be noted that we should tune
the K\"ahler moduli of vanishing two cycles in an appropriate way.
Note in particular that the moduli space of $K3$ does not
naturally split to K\"ahler and complex deformations.
This fact is going to be implicit in our discussions
of Calabi-Yau singularities below
and will not be mentioned further.

\newsec{Singular CY's, $c=1$ String and $U(n)$ Gauge Symmetry}

Let us summarize what we have learned from the example with
$K3\times T^2$.
The whole analysis of singularity was purely local.
We never used the global structure of $K3$ except for the
$N=4$ structure. The only relevant piece of geometry of $T^2$
was the existence of a circle $S^1$. Since the monodromy never
mixes the $A$ and $B$ cycles one can discuss them independently.

Now consider a family of 3-dimensional Calabi-Yau spaces.
It is natural to ask what happens if we have different types
of vanishing 3-cycles.    The 3-cycle can collapse to  {\bf A}) a
point,
{\bf B}) a line, or {\bf C}) a surface.  Case {\bf A} is encountered
in the conifold singularity of a Calabi-Yau represented locally by
\eqn\coni{x^2+y^2+z^2+w^2=\mu}
Note that the 3-cycle can be viewed as a real section of the above
(take $\mu$ to be real and positive) in which case it defines $S^3$.
  Note that as $\mu\rightarrow 0$ the whole sphere shrinks to a
point.
Compactifying Type IIB string on such Calabi-Yau one expects to end
up
with one hypermultiplet
of mass $\sim \mu$ \st.

For the example of  case  {\bf B}
one can  consider a family of $K3 \times T^2$, discussed above,
where $K3$ develops a singularity by shrinking  of  several 2-spheres
$S_i^2$ to zero size.

For examples of type {\bf C} where the vanishing cycle collapses
to a surface, the natural candidate is $T^2\times T^4$
compactification
where the modulus of $T^2$ goes to infinity--however this is
infinitely far away in the moduli of Calabi-Yau.  In fact
it is probably the case that the singularities of type
C do not occur at finite distance in the moduli of Calabi-Yau.

Locality leads us to a natural hypothesis that {\it whenever several
3-cycles shrink to
a circle $S^1$ and locally the geometry is ${\rm ALE} \times S^1
\times
{\bf R}$,
gauge symmetry appears and we find two $N=2$ multiplets,
a vector multiplet and a hypermultiplet both in the adjoint
representation of the gauge group.} This is the field content of $N=4$
Yang-Mills theory. In general we do not expect to have $N=4$
supersymmetry
for compactifications on
CY. Therefore we may expect to find other
$N=2$ hypermultiplets charged with respect to the gauge group which
break
$N=4$ down to $N=2$. Exactly this happens in the examples below.

Finally, it seems that all we need to obtain a nonabelian gauge symmetry
in
Type IIB compactification is a Calabi-Yau manifold with a singularity
along a complex curve $C$. In the vicinity of $C$ the Calabi-Yau
should
look like a fibration by ALE spaces. Also, the curve $C$ should have
at
least one noncontractible cycle (for example, $C={\bf C}^*\sim
S^1\times
{\bf R}$ is good enough).

Before embarking on examples of such singularities, let us recall
that
the conifold singularity is related to the $c=1$ string at the
self-dual
radius.
It is natural to wonder what type of singularity one obtains
when one considers $n$ times the self-dual radius.  The corresponding
ground ring has been computed \mukgh\
at $\mu =0$ to be (by a change of coordinates)
\eqn\ren{(x^2+y^2)^n+z^2+w^2=0}
This singularity has a structure similar to conifold and  \reso\
singularity,
namely if we take for instance $n=1$ it is of the same type as {\bf
A}
and if we take $\zeta=x^2+y^2$ it is of the same type as {\bf B}.

Let us consider the perturbation of \ren\ given by
\eqn\peren{W=\prod_{i=1}^{n}( x^2+y^2-\mu_i)+z^2+w^2=
\prod_{i=1}^{n}( \zeta -\mu_i)+z^2+w^2=0}
Note that the cosmological constant deformation of $c=1$
model at $n$-times the self-dual radius corresponds to all
$\mu_i=\mu$.
This more general deformation by arbitrary
$\mu_i$ seems to correspond to $n-1$ discrete states
with zero momentum (with the dressing, violating Seiberg's condition)
\ref\muket{D. Ghoshal, P. Lakdawala and S. Mukhi,
Mod. Phys. Lett {\bf A8}
(1993) 3187}.

For $\mu_i=\mu$ the manifold \peren\ is singular along the curve
\eqn\sinloc{(x^2+y^2-\mu)^{n-1}=0,~z=0,~w=0~.}
The transversal fiber is  the $A_{n-1}$ ALE space. Locally, around
the
singular locus the manifold looks like an ALE fibration over $
S^1\times
{\bf R}$.
We expect that this singularity corresponds to enhanced $SU(n)$
gauge
symmetry.
We will see below that $N=4$ is broken to $N=2$ by extra matter in
fundamental representation.

For generic $\mu_i$ it is easy to see that as long
as none of the $\mu_i=0$ and $\mu_i \not= \mu_j$ for all $i\not= j$
the non-compact threefold defined locally by \peren\ is non-singular.
The dimensions of $H^{1,1}$ and $H^{2,1}$ are given by $h^{1,1}=n-1$
and
$h^{2,1}=n$
which means that the corresponding type IIB theory will have $n-1$
hypermultiplets (in addition to the dilaton) and $n$ vector
multiplets.
Let us first understand why $h^{2,1}=n$:  The easiest way is to note
that
we have $n$ parameters $\mu_i$ which deform the theory away
from the singularity and they are in one to one correspondence
with elements of $H^{2,1}$.  Moreover we can give an explicit basis
for the corresponding
3-cycles as follows.  Consider $\mu_i\rightarrow 0$ and fix all
the other $\mu_j$'s at generic points.  Then it is easy to see by
examining
the defining equation
that we end up getting
a conifold singularity, i.e. a vanishing 3-cycle $S^3_i$.
In this way we can use $V_i=[S^3_i]$ to form a basis for the
vanishing
3-cycles.
Moreover according to \st\ for each $\mu_i \rightarrow 0$
we should get a massless hypermultiplet charged under the $U(1)$
whose scalar component corresponds to $\mu_i$.  To describe
$H^{1,1}$ note that we have a projection from our threefold
to a twofold given by sending $(x,y,z,w)\rightarrow (\zeta
=x^2+y^2,z,w)$
and we can pull back the corresponding $n-1$ elements of $h^{1,1}$
which are dual to the $n-1$ vanishing 2-cycles of $H_2$.  It is not
difficult to show that there are no other elements of $H^2$ and $H^3$
 with compact support.

\bigskip

\let\picnaturalsize=N
\def\picsize{2.0in}
\def\picfilename{ren2.eps}
\ifx\nopictures Y\else{\ifx\epsfloaded Y\else \fi
\global\let\epsfloaded=Y
\centerline{\ifx\picnaturalsize N\epsfxsize \picsize\fi
\epsfbox{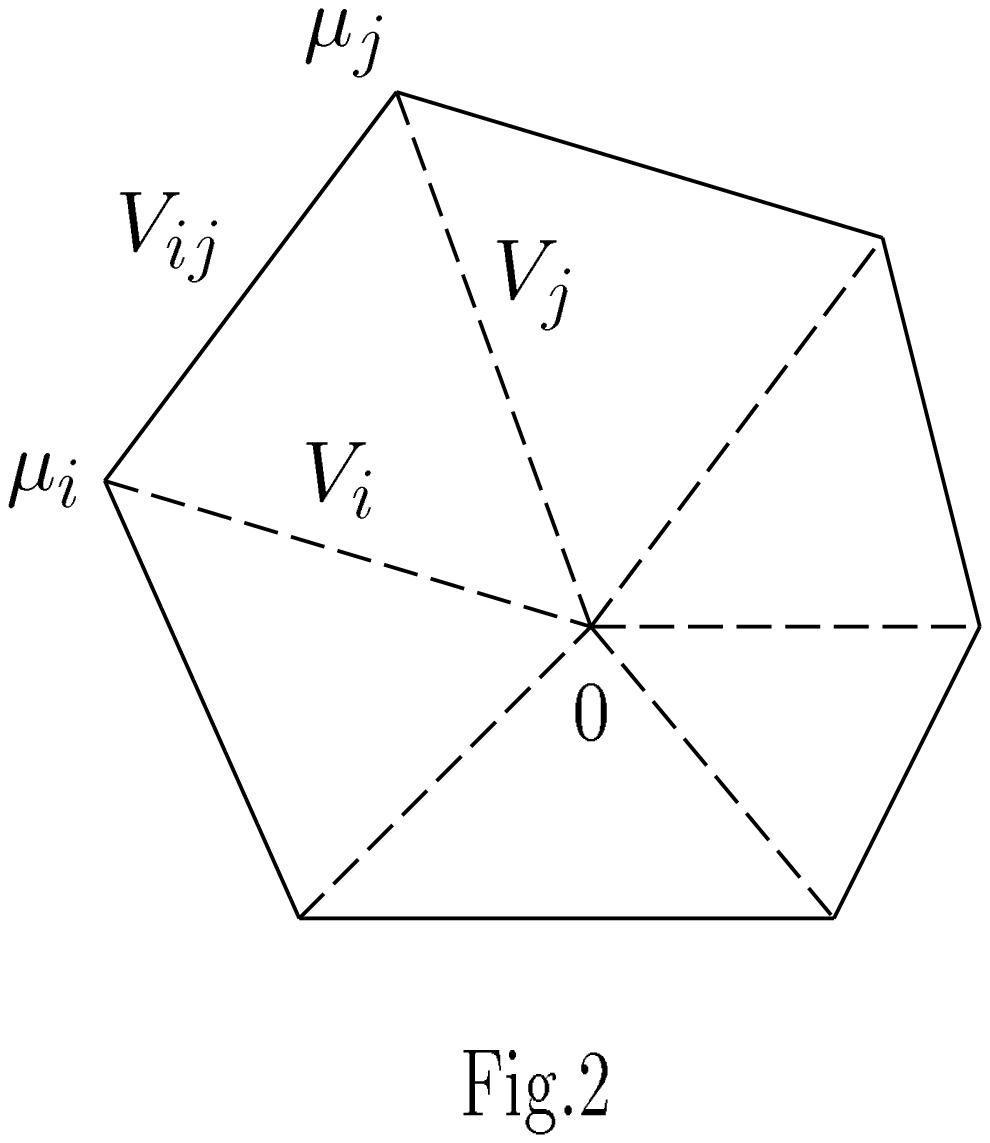}}}\fi

\bigskip

 Suppose we take generic $\mu_i$ except for
letting two of them approach each other.  Then not only we get a
vanishing
two cycle, but also we get a vanishing 3-cycle:  It suffices
to consider the case $n=2$, and take
$$W=(x^2+y^2-R^2)^2+z^2+w^2-r^2=0$$
with $R>>r$ real and positive.  Then $r$ measures the separation
of the two $\mu_i$ and $R$ measures their distance to the origin.
Then if we consider the above equation over the reals we see that
it has the topology of $S^2\times S^1$, where $r$ measures the radius
of $S^2$ and $R$ measures the radius of $S^1$ (more precisely
the radius of $S^1$ varies from $R+\sqrt r$ to $R-\sqrt r$).
This implies that as $r\rightarrow 0$ we have a vanishing
3-cycle of type B discussed above.
It is
easy to see that this 3-cycle is not independent of $V_i$
defined above.  In fact, if $\mu_i\rightarrow \mu_j$
the vanishing 3-cycle $V_{ij}$ which we realized as an $S^2\times
S^1$
is homologically equivalent to
$$V_{ij}=V_i-V_j$$
If the local analysis of the string duality is any hint, we can
expect that if two $\mu_i$ approach each other the $U(1)\times U(1)$
related to the two $\mu$'s get enhanced to
$U(1)\times SU(2)\sim U(2)$
vector multiplet with an extra adjoint hypermultiplet of $SU(2)$.
Moreover, there are also two massive hypermultiplets that form a
doublet of $U(2)$ (both of them become massless
as $\mu_i$ and $\mu_j$ approach zero).
The mass
is related to the fact the $\mu_i\not=0$ and thus we are in the
Coulomb
phase of the diagonal $U(1)$ which couples to the fields in the
fundamental
representation.
Note that this
is consistent with dynamics of gauge fields: in the infrared we can
ignore
massive fields, in which case the theory behaves as if it was an
$N=4$
theory
and we thus expect to have an exact $U(2)$ symmetry even in the
quantum
theory.  Note that this is needed in order for the above picture
to make sense because the singularity of the Calabi-Yau manifolds
must have a direct bearing on the exact, quantum corrected, physics.
In particular if we did not have the adjoint of the gauge group we
know
that
the $SU(2)$ symmetry disappears from the quantum moduli and we only
realize a $U(1)$ gauge symmetry in the infrared, in contradiction
with the
geometry of the Calabi-Yau.

To understand the full matter content of this theory consider
the action of the monodromy group on 3-cycles  $H_3$. For $A_{n-1}$
case
the
monodromy group coincides with  group $S_n$ of  permutation from $n$
elements.
It acts on cycles $V_{i j}$ and $V_i$ in an obvious way
by permuting the indices. Cycles $V_i$ correspond to the matter
sector. Assuming that there is a stable
3-brane configuration wrapped around one cycle  one immediately
concludes that  3-branes wrapped around all $V_i$ are stable solitons
(at least in some limit when all $\mu_i\rightarrow 0$). The set
of particles corresponding to $V_i$ should form the fundamental
representation of $U(n)$.

To summarize, we are suggesting that if $\mu_i=\mu, (i=1,... n)$ the
field
theory
dynamics is the same as for an $N=2$  $U(n)$ gauge theory  with
a massless adjoint hypermultiplet
of $SU(n)$ and a massive fundamental of $U(n)$ whose
mass is related to $\mu$.  As we move $\mu_i$ to generic
points the theory goes to the Coulomb phase
for which the surviving gauge group is $U(1)^n$.  Note also
that if we took into account the graviphoton the gauge
group is actually $U(n)\times U(1)$.

The parameters $\mu_i$ are related to the condensates of the scalar
field
$\phi$ of the vector multiplet as follows
${\rm det}(\phi - \zeta)= \prod (\mu_i- \zeta)$.
The fundamental field
is always massive. Of the adjoint hypermultiplet of $SU(n)$, $n-1$ of
them from Cartan remain massless and the rest
pick up mass.  Thus at generic $\mu_i$ the massless (field theory)
degrees of freedom are $n$ vector multiplets corresponding to
$U(1)^n$
and $n-1$ neutral hypermultiplets\foot{
It is also possible to give vev to
a cartan of the hypermultiplet. The gauge
symmetry is again broken to $U(1)^n$, but now
the fundamentals are still massless because
they don't couple to the hypermultiplet.}.  This is as we expected
because
$h^{1,1}=n-1,  h^{2,1}=n$.
 Since the infrared dynamics for $\mu \not =0$
is that of  $N=4$   $U(n)$ theory, the gauge coupling is not quantum
corrected.
This is in line with the fact that the prepotential in this theory
is expected to be
\eqn\prepp{F_0=\sum_i{1\over 2} \mu_i^2 {\rm log}\mu_i}
Note that if we set $\mu_i=\mu$ we get
$$F_0={n\over 2} \mu^2 {\rm log}\mu$$
This is consistent with the fact that at $R=n$ the modified coupling
constant $n\mu^2\rightarrow \mu^2$ is the one which does not
transform with $R\rightarrow 1/R$ duality, and is to be identified
with the $c=1$ cosmological constant.

Let us discuss a more complicated example with a gauge group
$U(n)\times U(m)$
and massive matter in a $(n,m)$ representation.  Consider  a
hypersurface
given by equation
\eqn\NM{\prod_{i=1}^n(x^2+y^2+\mu_i) + \prod_{j=1}^m(z^2+w^2+\nu_j)=M}
For $m=1$, this reduces to the previous case.  To see $U(n)\times U(m)$
symmetry,
take $\mu_i=\mu,\,i=1,\ldots ,n$,  $\nu_j=\nu,\,j=1,\ldots m$ and
$\mu^n=\nu^m=M$.  Then there is an  $A_{n-1}$ singularity along the curve
${\cal C}_1$: $ \{x^2+y^2+\mu=0,\,z=w=0\}$ and an
$A_{m-1}$ singularity along the curve ${\cal C}_2$: $
\{x=y=0,\,z^2+w^2+\nu=0\}$.  To see $nm$ 3-cycles corresponding to
a $(n,m)$
hypermultiplet, one takes $\mu_i \rightarrow 0,\, \nu_j \rightarrow
0$ for
every pair $(i,j)$. For  $(x,y,z,w)$ near origin,  \NM\ reduces to
a resolved
conifold:
$$ A(x^2 +y^2)+B(z^2+w^2)=M$$ which has a vanishing 3-sphere.  The
mass of the
corresponding 3-brane soliton is governed by $M$.

A compact version of this example can be given using double elliptic
fibrations over $\Pp^1$, discussed in  \ref\Hub{T.~Hubsch {\it Calabi-Yau
Manifolds}, World Scientific, 1991} and \ref\Sh{C.~Schoen, {\it J.~f\"ur
Math.}~{\bf 364} (1986) 85-111}. Namely, consider a complete intersection in
$\Pp^2\times\Pp^2\times\Pp^1$ given by two equations\eqn\man{\eqalign{
&P(x)z_1+Q(x)z_2 =0\cr
&R(y)z_1+S(y)z_2 =0\cr
}}
where $(z_1:z_2) $ are homogeneous coordinates of $\Pp^1$, $x$ and $y$ are
sets of coordinates of two $\Pp^2$'s and $P,Q,R,S$ are four homogeneous
polynomials of degree 3. Obviously, this complete intersection is a Calabi-Yau
three-fold, smooth for generic $P,Q,R,S$.

We can think of \man\ as a fibration over $\Pp^1$. Fixing a point
$(z_1:z_2)\in \Pp^1$ one sees that a fiber is a product of two tori --- cubics
in $\Pp^2$'s given by the first and the second equations in \man\
respectively. These two tori are completely independent. What 2-manifold does
one get if one takes a fibration over $\Pp^1$ by just one torus?

Examining this closer, one sees that an equation $P(x)z_1+Q(x)z_2=0$ defines
analmost del Pezzo surface --- a blowup of $\Pp^2$ in nine (special) points.
Indeed, when at least one of $P,Q$ is not $0$, we can find a single point in
$\Pp^1$ and when both $P=Q=0$ which happens in $3\times 3$ points on $\Pp^2$,
any $(z_1:z_2)$ solves the equation. This almost del Pezzo surface has
$h^{1,1}=10$ and  Euler characteristic 12. When we think of it as a toric
fibration over $\Pp^1$ it means that generically there are 12 singular fibers
where the torus degenerates. When two of these fibers collide, almost del
Pezzo surface develops a nodal $A_1$ singularity.

Let us return to the three-fold \man. To find its Hodge numbers we will employ
 two complimentary pictures where \man\ is a toric fibration over either of two
 almost del Pezzo surfaces given by the first or the second equations of
\man\ respectively. Taking pullback of $(1,1)$-forms from a surface to \man\
one gets 10 $(1,1)$-forms one of which is a K\"ahler class of $\Pp^1$. Thus
taking pullbacks from both surfaces one gets $10+9=19$ different
$(1,1)$-forms, so $h^{1,1}=19$. Since the Euler characteristic is zero,
$h^{2,1}=19$. All the complex deformations are the monomial deformations of
\man.

Now let us consider the picture of \man\ as a double toric fibration over
$\Pp^1$. On $\Pp^1$ there are 24 special points divided into two sets of 12
points. Over each  point from  the first (the second) set the first (the
second) torus degenerates. (It does not lead though, to a degeneration of
\man\ itself.) Changing coefficients in \man\ one can move two special points
on top of one another. Suppose {\it both} these points belong to {\it the
first set}, then there develops an $A_1$ singularity along the second torus.
Indeed, almost del Pezzo surface corresponding to fibration over $\Pp^1$ by
the first torus will have a node in this situation, and considering \man\ as a
fibration over that del Pezzo by the second torus one verifies the claim.

Now suppose that we move {\it one} point from {\it the first set} and {\it
another} from {\it the second set} on top of each other.
Then both tori degenerate over this point. It is not difficult to see that
this corresponds to a conifold singularity of the three-fold \man. Indeed,
near singularity each torus looks like a conic: $x_1 x_2=z$ and $y_1 y_2=z$,
where $z$ is a coordinate on $\Pp^1$. Combining these two equations one gets
the conifold equation:
$$x_1 x_2-y_1 y_2=0.$$

In general, $m$ points from the first set and $n$ points from the second set
coincide. Then there is a curve (torus) with $A_{m-1}$ singularity along it
and another curve, also a torus, with $A_{n-1}$ singularity.
Obviously, $m+n\leq 20$ and $0\leq m,n\leq 12$ in this example.

It will be very interesting to generalize the machinery discussed
above
for $D_n$ singularities which correspond to $SO(2n)$ gauge groups
\foot{The perturbed ground ring for $c=1$ is
given by
$W=y^2+(z+w^2-\mu)(z^{n-2}+x^2)$.}.

\subsec{D-brane Interpretation}
Given the fact that we found a dual D-string description
for the appearance of gauge fields for singular $ALE$ spaces
with $A_n$ singularity, it is natural to ask whether there
is any D-manifold
on which the light spectrum of D-string
 configuration matches the above spectrum.
We will first consider the singularity
corresponding to $n$-times the self-dual radius and
propose a corresponding $D$-manifold.  We will
then generalize it to the $U(n)\times U(m)$ considered
at the end of previous section.
Then we will show {\it why} the proposed $D$-manifold is
indeed the dual of the type IIB on the corresponding singular
Calabi-Yau threefold.

{}From the description given above it is clear that we need an additional
object corresponding to the point $\mu=0$ of Figure 2, and moreover
we expect
that the `lines' connecting $\mu_i$ to $0$ should now correspond
to hypermultiplets rather than vector multiplets.  Moreover we need
fewer supersymmetries, i.e., $N=2$ in 4 dimensions.  Thus
we should be looking for a D-manifold consisting of
$(n+1)$ $D$-brane skeletons, $n$ of which are of one type
and one is of a different type.  Consider the $n$ Dirichlet
branes of the first type,
which we call the $D_1$ type given by restricting
$$(x_6,x_7,x_8,x_9)=v_i$$
for $i=1,...,n$
where spacetime coordinates run from $x_0,...,x_9$.  The rest of the
coordinates satisfy
Neumann boundary conditions. The other D-brane
should not be parallel to this, otherwise we would get $N=4$
supersymmetry rather than $N=2$.  Moreover we want to get four
dimensional Poincare invariance.   So we consider the $D_2$ type to be
$$(x_4,x_5,x_6,x_7)=(0,0,0,0)$$
(with no loss of generality we have fixed the only $D_2$ brane
to have fixed boundaries at the origin of the coordinates).
The D-manifold consists of
 $n$ 5-branes of one type and one 5-brane of another
type which intersect on a $3+1$-dimensional spacetime
 if any of the $v_i=0$.
 It is
now
easy to see how we get the $SU(n)$ gauge symmetry and the adjoint matter:
They simply come from the open string sectors connecting the $D_1$ branes
to each other.  Note that the adjoint matter simply comes from
dimensional
reduction from 6 to 4, as we expected.  The $U(1)$ gauge symmetry comes
from the open string connecting $D_2$ to itself.  Now consider
open strings connecting any of the $D_1$ branes to the $D_2$ brane.
  It involves boundary conditions
on both sides which are DD, NN or DN (where N stands for Neumann and
D for Dirichlet).   Then the bosonic
oscillators are integral in the DD and NN directions but half-integral
in the DN directions.  Thus in the R  sector (NS sector) the NSR
fermions are
integral (half-integral) in the DD and NN directions but
half-integral in the
DN directions
(note that the supercurrent has integral (half-integral) expansion
in the R
sector
(NS sector)).
Looking at the two types of $D$-branes we see that in the light-cone
($x_0=\tau$, $x_1=\sigma $)
the directions  $x_2,x_3$ are NN type, $x_6,x_7$ are DD type
and $x_4,x_5,x_8,x_9$ are DN type.
We thus have the same structure as appears in the right-moving
twisted sector
of heterotic strings upon compactification on $K3=T^4/Z_2$.  We thus get
a system with $N=1$ in six dimensions or $N=2$ in four dimensions
which leads to one hypermultiplet  (to see that  is a hypermultiplet
it is sufficient to study the NS sector where
one finds four scalar degrees of freedom).
Taking into account the Chan-Paton factors at the open string ends
we see that
the
hypermultiplet is charged and that it is in the fundamental
representation
of $U(n)$  exactly as we were expecting!   Note also that the mass of
the hypermultiplet goes to zero if the $6,7$ components of the
corresponding
$v_i$ approach zero (i.e. we have to tune two parameters to get a
massless
state, as expected for the conifold).

\bigskip

\let\picnaturalsize=N
\def\picsize{2.0in}
\def\picfilename{ren3.eps}
\ifx\nopictures Y\else{\ifx\epsfloaded Y\else \fi
\global\let\epsfloaded=Y
\centerline{\ifx\picnaturalsize N\epsfxsize \picsize\fi
\epsfbox{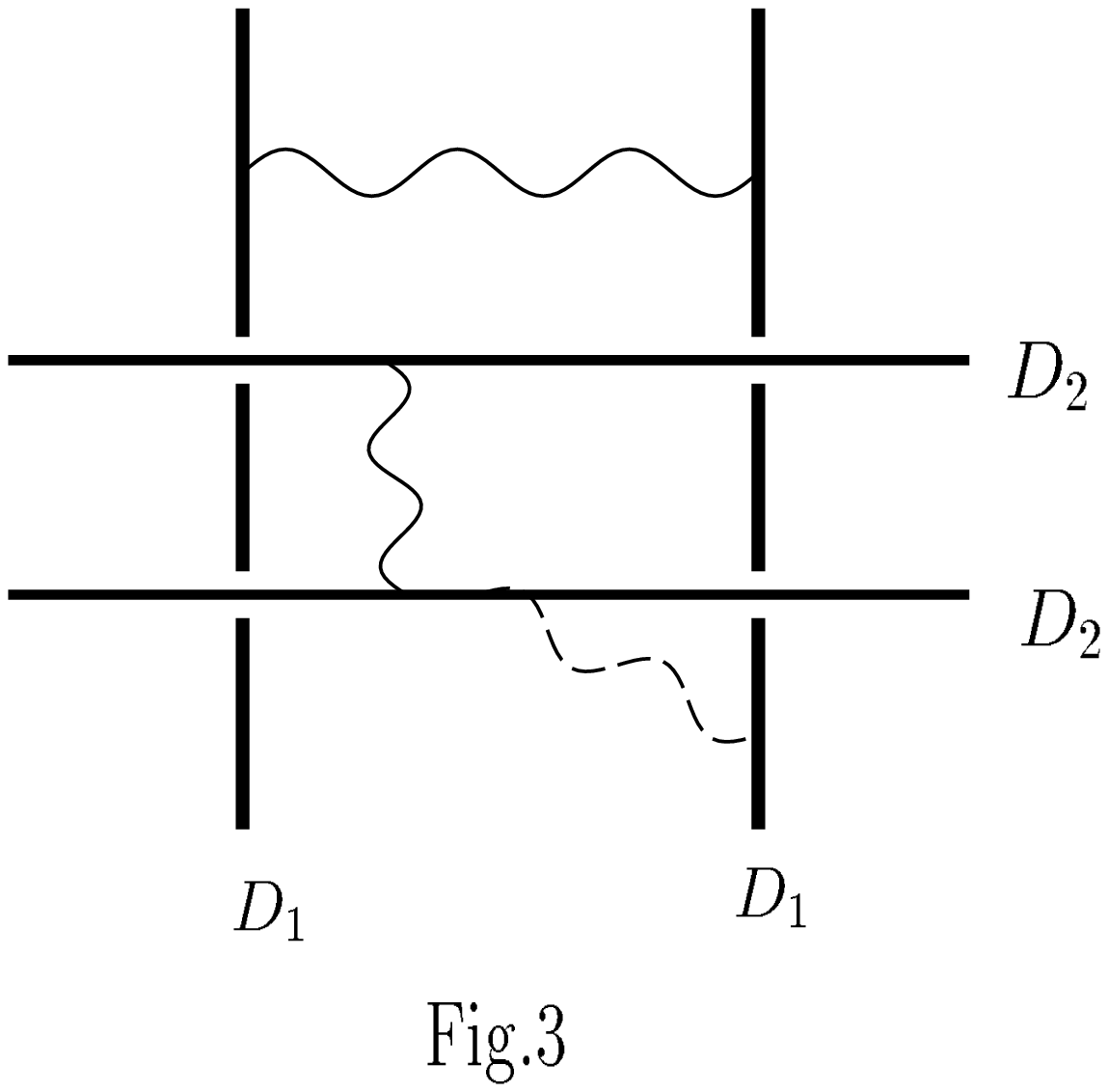}}}\fi

\bigskip

 It is now
clear how to generalize it to the
case of  $U(n)\times U(m)$ considered at the
end of last section.  In
that case we will naturally consider $n$ $D_1$ branes and $m$ $D_2$
branes.  It is clear that we will obtain the matter spectrum
in the $(n,m)$ representation
coming from the Chan-Paton factors lying on each of the two D-branes.

So far we have only proposed a D-string vacuum on a
D-manifold which leads to the same spectrum
as type IIB on
the corresponding singular Calabi-Yau.  Can we actually derive
this from the type IIB $SL(2,Z)$ duality in 10 dimensions just as we
did for gauge symmetry enhancement? Note in particular
that if we understand the case of $n=m=1$ the rest will follow
in an obvious way by `superposition'.  The case $n=m=1$
is the ordinary conifold singularity given by
$$xy-zw=\mu$$
Let us introduce another variable $\zeta$ and write the conifold
equivalently as
$$xy=\zeta $$
$$zw=\zeta -\mu$$
The H-version of the `stringy cosmic string' picture thus gives us
one $H$ string, whose projection is at $\zeta =0$ and another one
whose projection is at $\zeta =\mu$.  The corresponding fibers
$xy={\rm const}$ and
$zw={\rm const}$ are independent of each other.  Thus applying
the duality argument for each one, sketched before we obtain
two 5-branes, which we can easily identify with $D_1$ and $D_2$
described above!  Note that in this correspondence we had
to apply $R\rightarrow 1/R$ duality twice, once
for each fiber; thus type IIB
on the conifold is equivalent to type IIB again, but now
with suitable $H$-fields turned on.  Applying $R\rightarrow 1/R$
twice has shifted the value of $p=3$
for the solitonic $p$-brane by two units down to $p=1$,
and the soliton is now realized
as an elementary $D$-string state.

Note in particular
that we have shown that the existence and the degeneracy
of the solitonic massless hypermultiplet
for type IIB strings on a conifold is a consequence of the conjectured
$SL(2,{\bf Z})$ duality of
type IIB in 10 dimensions and is in agreement with the
conjecture of Strominger \st .

\newsec{Final Remarks and Conclusions.}

We have seen that the appearance of massless states
in type II compactifications follows from
$SL(2,Z)$ duality of type IIB in ten dimensions.
It can be described  perturbatively using D-strings propagating
on suitable D-manifolds.  Moreover
they are dual to type IIB in the presence of D-branes and the
massless states (be they gauge bosons or massless
charged matter) can be viewed as light states in
an open string subsector of the D-string in the presence of D-branes.

We described the appearance of $A_n$ gauge symmetry
through Chan-Paton factors of the open string sector of D-string.
We could have talked about the $D$ and $E$ series, in which case
what is most likely to happen is that not all the configurations of open
strings between the D-branes are stable (similarly to the story of
solitons of
$N=2$ Landau-Ginzburg theories in 2 dimensions \ref\wav{P. Fendley, S.D.
Mathur, C. Vafa and N.P. Warner, Phys. Lett. 243{\bf B} (1990) 257}),
and the stable
ones reproduce the expected root lattice of the group.  This deserves
further study.

So far we have talked about non-compact D-manifolds.
This was because that was sufficient to describe
the dual to the local singularity of Calabi-Yau. However
if we are to consider the dual to the compact Calabi-Yau,
the D-manifold will have to be compact.  Once we do so
we have to develop the notion of D-manifolds rather carefully.
There can be various type of restrictions.  In particular
if we take the original type IIB compactification with $H$-field
turned on, the condition of conformal invariance and the value
of central charge puts restrictions.  Thus taking its dual we
will have dual restrictions for the D-manifold.
Say if we begin compactifying the space on a two dimensional torus,
and consider $D$ 5-branes whose Dirichlet boundaries partially
reside on the torus
the story is going to be changed.
In fact by considering its dual
which is the $H$ dual of the `stringy cosmic strings', the
restriction on the
number
of cosmic strings being less than
24 \cosm\ puts a bound on the number of 5-branes
(note also that this implies that their positions are not
independent --- a key
point
which has to be studied further).

The relation between singularities of Calabi-Yau threefolds and
non-critical
string theory
is still quite mysterious.  In particular it would be important
to explain the partition function of non-critical strings
in terms of physical spectrum of Calabi-Yau compactification.  For
example
it would be interesting to check  the
correction to
$F_1$ which is expected
to be
$$F_1={-1\over 24}(n+{1\over n}){\rm log}\mu$$
and compute the correction to $R^2$ \bcov
\narainet.
Note that unlike the case considered in
\ref\vb{C. Vafa, Nucl. Phys. {\bf B447} (1995) 252}\
as $\mu \rightarrow
0$,
we have a strongly interacting
$U(n)$ theory, for which the computation of $R^2$ may require
understanding some subtleties.
Also of interest would be to study realizations of the $R=n$
singularities
in compact Calabi-Yau manifolds.
These issues are currently under investigation\foot{
A promising candidate for this is the strong coupling
locus $y=1$ appearing
in the examples of $N=2$ duality considered in
\ref\kv{S. Kachru and C. Vafa, Nucl. Phys. {\bf B450}  (1995) 69}.
This is currently being investigated \ref\mor{D. Morrison,
private communication.}\ with results similar to what one
expects from the picture presented here.}

Another interesting direction to pursue is the duality
betweeen type IIA on
$K3$ with type I on $T^4$.  It is likely that the D-string
formulation of type IIA on $K3$ is very close to the type I
compactification
on $T^4$ and it is likely to be a useful set up in understanding this
duality better.

Finally we would like to make a remark in connection with
the stringy realization of dualities in the $N=1$ supersymmetric
gauge theories (see \ref\ints{K. Intriligator and N. Seiberg, {\it
Lectures on
Supersymmetric Gauge Theories and Electric-Magnetic Duality}, Preprint
RU-95-48,
hep-th/9509066}\
for a review).  One of the simplest examples, which is accidentally
an $N=2$ system, is an electric $SO(3)$ gauge theory with a triplet
matter
which is dual to a magnetic theory of $SO(2)$ with a doublet and
a neutral meson.  This is in fact the same as the $N=2$ duality
of Seiberg and Witten
\ref\sw{N. Seiberg and E. Witten, Nucl. Phys. {\bf B426}  (1994) 19},
where the electric theory
is equivalent to $N=2$ theory of $SU(2)$ Yang-Mills and the magnetic
theory is equivalent to the $U(1)$ theory with a massless hypermultiplet
corresponding to the massless magnetic monopole of the $SU(2)$
electric theory.
  One may at first think that this duality
has already been discovered in string theory \kv\
where the electric side is the heterotic one and the magnetic side
is the type II side.
However that would not be quite accurate:  the magnetic side contains
{\it perturbative} states which are charged under the $SO(2)$
whereas in the type II case the charged hypermultiplets
are still 3-brane solitonic states!  What we have found here is that
we should expect the magnetic side in the string theory to be realized
as a D-string on a D-manifold where the massless $U(1)$ charged
states are given by perturbative states.  One should try to construct
the D-manifold equivalents of {\it compact}
Calabi-Yau compactifications to
realize this idea more concretely.  In fact since the examples
of \kv\ are $K3$ fibrations \ref\leret{A. Klemm, W. Lerche and P. Mayr,
Phys. Lett. 357{\bf B} (1995) 313}\ the
dual D-manifold can easily be described as a similar fibration
with the fibers replaced by the appropriate configuration of D 5-branes.

Since we have now realized a duality in string theory
in the same sense as duality in supersymmetric field theories
 one may hope to
obtain more general classes by studying D-strings on D-manifolds
which leave only $N=1$ supersymmetry
in four dimensions.   It is likely that the duality
between D-strings and heterotic strings in the $N=2$ case
generalizes to interesting $N=1$ dualities.

We would like to thank A.~Dabholkar, D.~Ghoshal,
J.~Harvey, S.~Kachru, E.~Martinec, S.~Mathur, D.~Morrison, H.~Ooguri,
T.~Pantev,
S.~Shenker, E.~Witten and
S.-T.~Yau for valuable discussions.

The research of  V.~S.~is supported in part by NSF grant DMS 9304580
The research
of M.~B.~and C.~V.~is supported in part by NSF grants PHY-92-18167 and
PHY-89-57162.  The research of M. B. was also supported by NSF 1994
NYI award
and DOE 1994 OJI award.

\listrefs
\end